\documentstyle[epsfig,12pt]{article}

\def\pl{Phys.\ Lett.}
\def\prl{Phys.\ Rev.\ Lett.}
\def\np{Nucl.\ Phys.}

\def\prv{Phys.\ Rev.}

\def\ijmp{Int.\ J.\ Mod.\ Phys.}
\def\>{\rangle}
\def\<{\langle}

\begin{document}

\pagestyle{empty}
\begin{flushright}
CERN-TH.6990/93\\
UA/NPPS-5-93
\end{flushright}
\vskip 60pt
\begin{center}
{\Large\bf{Criticality, Fractality and Intermittency
in \\
\medskip
Strong Interactions}}
\vskip 45pt
\large{N.G. Antoniou, F.K. Diakonos, I.S. Mistakidis}
\vskip 5pt
{\large\it
{Department of Physics, University of Athens,\\
\smallskip
GR-15771, Athens, Greece}}
\vskip 12pt
\large{and}
\vskip 12pt
\large{C.G. Papadopoulos}
\vskip 5pt
{\large\it {CERN, TH Division, CH-1211, Geneva 23, Switzerland}}
\end{center}
\vskip 80pt
\begin{center}
{\large\bf{Abstract}}
\end{center}
\medskip
Assuming a second-order phase transition for the hadronization process,
we attempt to associate intermittency patterns in high-energy hadronic
collisions to fractal structures in configuration space and
corresponding intermittency indices to the isothermal critical
exponent at the transition temperature. In this approach,
 the most general multidimensional intermittency pattern, associated
to a second-order phase transition of the strongly interacting system,
 is determined, and its relevance to present and future experiments is
 discussed.
\vfill
\begin{flushleft} CERN-TH.6990/93\\August 1993\\
\vfill
\end{flushleft}
\vfill

\newpage
\pagestyle{plain}
\setcounter{page}{1}

\begin{center}
\section{Introduction}
\end{center}
\par One of the central issues in quark-gluon plasma physics is related to
the nature and structure of the quark-hadron phase transition and its
measurable effects on the pattern of multihadronic states generated in
a high energy-density collision of relativistic ions. Recent studies in
lattice gauge theories suggest that the QCD phase transition for
two flavours $(f=2)$ and massless quarks ($m_{u}=m_{d}=0$) is of second
order and this result persists even for $f=3$ provided that the
mass of the strange quark is sufficiently large
($m_{s} \geq 100 \; \mbox{MeV}$) \cite{hats:1}.
This development brings as to
the hypothesis of a second-order chiral phase
transition in QCD close to the real world and suggests a further
exploitation of this fundamental assumption in an attempt
to build-up a self-consistent dynamical scheme of strong interactions
at the transition temperature ($T=T_{c}$). In a recent work
along this line F.Wilczek has argued that the hypothetical
second-order chiral phase transition of the $f=2$ QCD belongs to the
universality class of the $n=4$ Heisenberg magnet; therefore, in
order to study the behaviour of the strongly interacting system
near the critical point, one may employ the critical exponents of
this magnetic system as given by the fixed point of the corresponding
renormalization group \cite{wilz:1}. In particular, the power law
at $T=T_{c}$ connecting the condensate $\<{\bar {q}}q\>$
(order parameter) with the
quark mass (ordering field), $\<{\bar{q}}q\> \sim m^{1/ \delta}$,
in the limit $m \rightarrow 0$ ($m=m_u=m_d$)
introduces a universal critical exponent $\delta$ in the
strongly interacting system, which reflects the non-perturbative sector
of the underlying theory.
A similar power law, $\<\rho\> \sim m_{\pi}^{1/ \delta}$, with
the same exponent $\delta$, relating
the density of hadrons (pions) in configuration space, $\<\rho\>$, with
the pion mass $m_{\pi}$ (chemical potential) in the chiral limit
($m_{\pi} \rightarrow 0$), is expected to hold in the hadronic phase
of a quark-gluon production process, at $T=T_{c}$. The fundamental index
$\delta$ has a distinct geometrical interpretation giving the fractal
dimension,
$D_{F}= {\delta D \over 1+ \delta}$, of the
critical system at $T=T_{c}$, embedded in a support space of
dimension $D$ \cite{stin:1}.
\par In this work, adopting the hypothesis of a second-order
phase transition in the strongly interacting system, we attempt to
incorporate the above geometrical property of the critical system
into the hadronic $S$-matrix for multiparticle production
and show how the fractal structure of the critical system in configuration
space may induce multidimensional intermittency patterns in
momentum space, in accordance with experimental evidence.
For this purpose we derive a suitable integral representation
for multiparticle densities in momentum space, introducing proper
light-cone variables in the treatment of the reduced $S$-matrix
elements. In this approach the constraint of self-similarity
(monofractality) in configuration space, required by the hypothesis
of a second-order phase transition, is easily incorporated and its effect
on the particle densities in momentum space is extracted.
As a result, the general fractal solution leading to a complete
multidimensional intermittency pattern in momentum space is obtained,
expressed as a set of power laws for the factorial moments of all orders.
This pattern depends on two universal indices, one of which is
the critical
exponent $\delta$. Therefore, a systematic comparison with
experiments may
give, not only an indication for the onset of a second-order phase
transition in the quark-gluon system, but also a measurement of the
corresponding critical exponent at $T=T_c$.
\par In section 2 the formalism leading to a complete set of hadronic
densities in $3+1$ space-time is presented by considering the inclusive
$S$-matrix treatment of a hadronic collision within the framework
of inside-outside cascade dynamics \cite{bjor:1}.
In section 3, the geometrical
constraint of fractality as a requirement of a second-order QCD phase
transition at $T=T_c$ is imposed and the associated fractal solution
in $3D$-momentum space is found and discussed. In section 4, the resulting
multidimensional intermittency pattern showing strong fluctuations in
small domains of momentum space, is derived. This pattern, combined
with a smooth background representing conventional correlations in
momentum space \cite{curr:1},
is compared, in section 5, with the typical behaviour
of the factorial moments observed in high-energy hadronic collisions.
Finally, in section 6 our results are summarized and discussed and
a critical comparison with other approaches is attempted.
\begin{center}
\section{Hadronic densities in $3+1$ space-time}
\setcounter{equation}{0}
\end{center}
\par Consider the inclusive process
\[h_A+h_B \rightarrow h_1 + h_2
+...+h_p+ \mbox{anything}\]
in the hadronization sector of the collision,
defined as the component of the $S$-matrix connecting the initial hadrons
$(h_A, h_B)$ with final states consisting of newly hadronized particles
(at $T=T_c$) in the process of a second-order QCD phase transition.
The densities ${\tilde{\rho}}_c({\vec{q}}_1,..., {\vec{q}}_p)$
in momentum space are written as follows:
\begin{equation}
{\tilde{\rho}}_c({\vec{q}}_1,..., {\vec{q}}_p)={E_1 E_2 ... E_p \over
\sigma_h} { d \sigma^{(p)}_{in} \over d^3 {\vec{q}}_1 ...d^3 {\vec{q}}_p}
\sim \sum_{\{k_a\}} \vert
\<q_1 ... q_p; \{ k_a \} \; \mbox{out} | p_A p_B
\; \mbox{in}\> \vert^2
\end{equation}
where $q_i=({\vec{q}}_i,E_i)$ are the four-momenta of hadrons (pions)
$h_1 ... h_p$ in the final state, $\sigma_h$ the total cross section
of the
collision $h_A+h_B$ in the hadronization sector,
and $d \sigma^{(p)}_{in}$
the $p$-hadron inclusive cross section in the same sector.
The summation runs over the hadronic states that build-up the missing
mass in the inclusive process with a total four-momentum
$k=\sum_a k_a$.
The matrix elements in eq.(2.1) connect the initial state
$|p_A p_B ; \mbox{in}\>$ with final states $|\mbox{out}\>$
generated at $T=T_c$
by the quark-hadron phase transition
(the critical component of the $S$-matrix).
Reducing now the $S$-matrix
elements in eq.(2.1) we remove the hadrons $h_1...h_p$ from the outgoing
states $|\mbox{out}\>$ and
using completeness of the states $|\{k_a\}\>$ we
finally obtain \cite{bjor:1,muel:1}:
\begin{equation}
{\tilde{\rho}}_c({\vec{q}}_1,..., {\vec{q}}_p)=f_{AB} \int \prod_{i=2}^p
d^4 x_1 d^4 x_i d^4 z_i e^{-i q_1 x_1} e^{-i q_i (x_i - z_i)}
G(x_1...x_p;z_2...z_p)
\end{equation}
\begin{eqnarray}
G(x_1...x_p;z_2...z_p)&=&K_{x_1}...K_{x_p}K_{z_2}...K_{z_p}
\nonumber \\ &\cdot& \<p_A p_B|T \{ \phi(x_1)...\phi(x_p) \} T
\{ J(0) \phi(z_2)...\phi(z_p) \}|p_A p_B\>
\end{eqnarray}
where $f_{AB}=(2 \pi)^5/(2^{p+2} W P\sigma_h)$ is a
normalization factor corresponding to a total energy $W$ and incident
momentum $P=|{\vec{P}}_A|=|{\vec{P}}_B|$ in the centre-of-mass system.
The
current $J(0)$ in eq.(2.3) is associated with the pion field $\phi(x)$
and $K_x$ denotes the corresponding Klein-Gordon operator. Changing
integration variables in eq.(2.2): $x_i^{(+)}=x_i+z_i$
$r_i=x_i-z_i$ $(i=2,...,p)$ and putting $x_1 \equiv r_1$, we have:
\begin{equation}
{\tilde{\rho}}_c({\vec{q}}_1,..., {\vec{q}}_p)={f_{AB} \over 2^{p-1}}
\int \prod_{i=2}^p
d^4 r_1 d^4 r_i d^4 x_i^{(+)}
e^{-i q_1 r_1} e^{-i q_i r_i} {\cal{G}}(r_1..r_p;x_2^{(+)}...x^{(+)}_p)
\end{equation}
Integrating over the variables $x_i^{(+)}$ we finally obtain:
\begin{eqnarray}
{\tilde{\rho}}_c({\vec{q}}_1,..., {\vec{q}}_p)&=& \int
d^4 r_1 ... d^4 r_p \exp(-i q_1 r_1 - ... - i q_p r_p) \rho(r_1,..,r_p)
\nonumber \\
\rho(r_1,...,r_p)&=&{f_{AB} \over 2^{p-1}} \int
\prod_{i=2}^p d^4 x_i^{(+)}
{\cal{G}}(r_1...r_p;x_2^{(+)}...x^{(+)}_p)
\end{eqnarray}
In the hadronization sector the space-time development of the collision
is
assumed to follow inside-outside cascade dynamics; it is therefore
suitable to use, in the longitudinal space-time $(r_{\parallel},t)$,
 light-cone
variables: $\tau=(t^2-{r_{\parallel}}^2)^{1/2},
\xi={1 \over 2} \ln \vert
{r_{\parallel}+t \over r_{\parallel}-t} \vert$.
With these new coordinates we have:
\begin{eqnarray}
q \cdot r = E t - {\vec{q}} \cdot {\vec{r}}&=&
{1 \over 2}(E+q_{\parallel})(t-r_{\parallel})
+{1 \over 2}(E-q_{\parallel})(t+r_{\parallel})
-{\vec{q}}_{\perp} \cdot {\vec{r}}_{\perp}
\nonumber \\
\exp(iqr)&=& \exp(-i {\vec{q}}_{\perp} \cdot {\vec{r}}_{\perp})
\exp[i m_T \tau \cosh (\xi-y)]
\end{eqnarray}
where $m_T$ (transverse mass) and $y$ (rapidity) are given by the
well-known relations~: $m_T=(q_{\perp}^2 + m_{\pi}^2)^{1/2}$,
$y={1 \over 2} \ln \vert {E+q_{\parallel} \over E-q_{\parallel}} \vert$.
The Fourier integral (2.5) is now written as follows:
\begin{equation}
{\tilde{\rho}}_c({\vec{q}}_1,..., {\vec{q}}_p)=\int \prod_{i=1}^p
\tau_i d \tau_i d \xi_i d^2 {\vec{r}}_{\perp i}
\exp[i {\vec{q}}_{\perp i} \cdot {\vec{r}}_{\perp i} - i \tau_i m_{T i}
\cosh(y_i-\xi_i)] \rho(r_1,...,r_p)
\end{equation}
where integration over the
impact parameter space (${\vec{r}}_{\perp}$) has
been explicitly introduced.
The physical picture we have adopted in this approach (inside-outside
cascade mechanism and a
second-order phase transition in thermal equilibrium)
requires strong constraints on the integrand $\rho(r_1,...,r_p)$
in eq.(2.7).
In fact since the development of the strongly interacting system in the
longitudinal space-time is assumed to take place along the hyperbolas
$t^2-r_{\parallel}^2 = \tau^2$ and
the narrow bands $\xi_i \approx y_i$ (inside-outside cascade), the support
of the functions $\rho(r_1,...,r_p)$ in $1+1$ space-time $(\xi, \tau)$ is
restricted as follows:
\begin{equation}
\rho(r_1,...,r_p) \sim \delta(\xi_p-y_p) \prod_{i=1}^{p-1} \delta(\xi_i-y_i)
\delta(\tau_i-\tau_p)
\end{equation}
With this constraint and after integrating over proper time variables
$\tau_i$, eq.(2.7) is simplified as follows:
\begin{equation}
{\tilde{\rho}}_c(y_i,{\vec{q}}_{\perp i})=
\int \{ d^2 {\vec{r}}_{\perp i} \}
\exp \Bigl(i \sum_{i=1}^p {\vec{q}}_{\perp i}
\cdot {\vec{r}}_{\perp i}\Bigr) \rho_c(y_i,{\vec{r}}_{\perp i};M_T)
\end{equation}
where ${\tilde{\rho}}_c(y_i,{\vec{q}}_{\perp i}) \equiv
{\tilde{\rho}}_c({\vec{q}}_1,..., {\vec{q}}_p)$,
$\rho_c(y_i,{\vec{r}}_{\perp i};M_T) \equiv
\rho_c(y_1...y_p,{\vec{r}}_{\perp 1}...{\vec{r}}_{\perp p};M_T)$,
$\{d^2 {\vec{r}}_{\perp i} \} \equiv
d^2 {\vec{r}}_{\perp 1}...d^2{\vec{r}}_{\perp p}$ and
$M_T=\sum_{i=1}^p m_{T i}$.
Imposing thermal behaviour on the
transverse mass distribution, we may write
$\rho_c(y_i,{\vec{r}}_{\perp i};M_T)=\exp(-{M_T \over T_c})
\rho_c(y_i,{\vec{r}}_{\perp i})$, and since the Boltzmann factor does not
affect the intermittency properties in momentum space we absorb it in
$\rho_c(y_i,{\vec{q}}_{\perp i})$ and finally obtain:
\begin{equation}
{\tilde{\rho}}_c(y_i,{\vec{q}}_{\perp i})=
\int \{ d^2 {\vec{r}}_{\perp i} \}
\exp\Bigl(i \sum_{i=1}^p {\vec{q}}_{\perp i}
\cdot {\vec{r}}_{\perp i}\Bigr) \rho_c(y_i,{\vec{r}}_{\perp i})
\end{equation}
where $\rho_c(y_i,{\vec{r}}_{\perp i})$ represents hadronic densities in
configuration space at $T=T_c$.
\par We consider the integral representation (2.10) as a model equation, based
on rather general assumptions. It connects, at $T=T_c$, observable density
fluctuations in $3D$ momentum space $(y,{\vec{q}}_{\perp})$ with corresponding
density fluctuations in $3D$ configuration space $(y,{\vec{r}}_{\perp})$.
It is, therefore, a suitable equation for studying multidimensional
intermittency patterns associated to self-similar structures of the hadronic
system in real space, at $T=T_c$.
\begin{center}
\section{Criticality and Fractality}
\setcounter{equation}{0}
\end{center}
Our basic hypothesis in this treatment is a simplified picture for the
hadronization process, in a high energy-density collision, as a static,
second-order QCD phase transition in thermal equilibrium. It generates
at $T=T_c$ strong density fluctuations resulting in the development of
a fractal dimension $D_F$ in real space according to the standard behaviour
of thermal systems undergoing such a transition and characterized by an
isothermal critical exponent $\delta$ at $T=T_c$ $(D_F={\delta D \over
1 + \delta})$ \cite{stin:1}.
In this picture it is natural to assume that the
hypothetical second-order QCD phase transition manifests itself as a critical
behaviour of the classical $3D$ system (hadronic fluid) specified at $T=T_c$
by the complete set of densities
$\{ \rho_c(y_1...y_p,{\vec{r}}_{\perp 1}...{\vec{r}}_{\perp p}), \;
p=1,2,... \}$
introduced in eq.(2.10). With this interpretation, the characteristic
power law of the chiral phase transition at $T=T_c$, $\<{\bar {q}}q\> \sim
m^{1/ \delta}$, valid in the quark phase ($m$ is the quark mass), must
have its counterpart in the hadronic phase, with the same critical exponent
$\delta$:
\begin{equation}
\<n\> \sim \mu^{1/ \delta} \; \; \; \; \; \;
\<n\>={\partial \ln Q_c \over
\partial \mu} \; \; \; \; (T=T_c)
\end{equation}
where $\<n\>$ is the average multiplicity of hadrons, $Q_c$ the grand partition
function at $T=T_c$, given in terms of the densities
$\rho_c(y_i,{\vec{r}}_{\perp i})$, and $\mu$ the chemical potential for the
production process, normally identified with the pion mass $(\mu=m_{\pi})$.
\par For completeness, one can easily verify that the critical behaviour (3.1)
of a classical system in $D$ dimensions is associated to a self-similar
structure in configuration space corresponding to a fractal dimension
$D_F={\delta D \over
1 + \delta}$. For this purpose, consider the grand partition function
$Q_c(\mu,V)$ of such a system, at $T=T_c$, in terms of the densities
$\rho_c({\vec{x}}_1,...,{\vec{x}}_p)$ $(p=1,2...)$ as follows:
\begin{equation}
Q_c(\mu,V)=\sum_p {(z-1)^p \over p!} \int_V \{d^D {\vec{x}}_i \}
\rho_c({\vec{x_1}},...,{\vec{x}}_p)
\end{equation}
where $z=\exp(\mu/ T_c)$ and $V$ is the volume of the system in $D$ dimensions.
A self-similar structure in configuration space with fractal
dimension $D_F$ requires:
\begin{equation}
\rho_c(\lambda {\vec{x}}_1,..., \lambda {\vec{x}}_p)=\lambda^{(D_F-D)p}
\rho_c({\vec{x}}_1, ...,{\vec{x}}_p)
\end{equation}
and in the limit $\mu \rightarrow 0$ eqs.(3.2) and (3.3) lead to a scaling
property, $Q_c(\mu,V)=F(\mu V^{D_F/D})$, as follows:
\begin{equation}
Q_c(\mu,V)=\sum_p \biggl({\mu V^{D_F/D} \over T_c}\biggr)^p {1 \over p!}
\int_{|{\vec{\xi}}_i| \leq 1} \{ d^D {\vec{\xi}}_i \}
\rho_c({\vec{\xi}}_1,...,{\vec{\xi}}_p)
\end{equation}
In the thermodynamic limit $(V \rightarrow \infty)$ the scaling form
$Q_c=F(\mu V^{D_F/D})$ combined with the normal boundary  condition
$Q_c \sim \exp[f(\mu)V]$, leads to the critical behaviour (3.1) with
$\delta={D_F \over D - D_F }$, giving finally $D_F={\delta D \over
1 + \delta}$. This simple relation of the critical exponent $\delta$
with the fractal dimension $D_F$ of the system at $T=T_c$ together with
eq.(2.10) provides a basic ingredient in our approach since it makes
the connection of the hadronic intermittency with a second-order phase
transition \cite{satz:1}
more transparent and easier to understand.
\par In this context and in order to establish a link between intermittency
patterns in momentum space with criticality of the hadronic system at $T=T_c$
one has to search, using eq.(2.10), for the most general class of fractals
in momentum space $(y,{\vec{q}}_{\perp})$ generated by a self-similar structure
in $3D$ configuration space $(y,{\vec{r}}_{\perp})$ corresponding to a fractal
dimension $D_3={3 \delta \over 1 + \delta}$. For this purpose and in order
to impose self-similarity, we introduce new coordinates in both spaces,
appropriate to the fact that the system, being finite, is not necessarily
translational invariant. We have:
\begin{eqnarray}
{\vec{r}}_{\perp i} \; ' &=& {\vec{r}}_{\perp i} - {\vec{R}}_{\perp}
\; \; \; \; \; \; (i=1,...p-1) \; \; \; \; \; \;
{\vec{R}}_{\perp}={1 \over p}({\vec{r}}_{\perp 1}+...+{\vec{r}}_{\perp p})
\nonumber \\
{\vec{q}}_{\perp i} \; ' &=& {\vec{q}}_{\perp i}-{\vec{Q}}_{\perp i}
\; \; \; \; \; (i=1,...p-1) \; \; \; \; \; \;
{\vec{Q}}_{\perp}={1 \over p}({\vec{q}}_{\perp 1}+...+{\vec{q}}_{\perp p})
\nonumber \\
y_i ' &=& y_i-Y \; \; \; \; \; \; \; \; \; \; (i=1,...p-1)
\; \; \; \; \; \;\;
Y={1 \over p}(y_1+...+y_p)
\end{eqnarray}
With these new variables, the problem of finding fractals in momentum space,
associated to self-similar structures in configuration space is formulated
as follows~:
\begin{eqnarray}
{\tilde{\rho}}_c(y_i ', {\vec{q}}_{\perp i} \; ';Y,{\vec{Q}}_{\perp})&=&
\int d^2 {\vec{R}}_{\perp} e^{i {\vec{Q}}_{\perp} \cdot {\vec{R}}_{\perp}}
\int \{ d^2 {\vec{r}}_{\perp i} \; ' \} \nonumber \\
&\cdot& \exp\Bigl(i \sum_{\mu,\nu}^{p-1}
\alpha_{\mu \nu} {\vec{q}}_{\perp \mu} \; ' {\vec{r}}_{\perp \nu} \; '
\Bigr)
\rho_c(y_i ',{\vec{r}}_{\perp i} \; ';Y,{\vec{R}}_{\perp})
\end{eqnarray}
\begin{equation}
\rho_c(\lambda y_i ', \lambda {\vec{r}}_{\perp i} \; ';
Y,{\vec{R}}_{\perp})= \lambda^{(D_3-3)(p-1)}
\rho_c(y_i ',{\vec{r}}_{\perp i} \; ';Y,{\vec{R}}_{\perp})
\end{equation}
\begin{equation}
{\tilde{\rho}}_c(\lambda y_i ', \lambda {\vec{q}}_{\perp i} \; ';
Y,{\vec{Q}}_{\perp})= \lambda^{({\tilde{D}}_3-3)(p-1)}
{\tilde{\rho}}_c(y_i ',{\vec{q}}_{\perp i} \; ';Y,{\vec{Q}}_{\perp})
\end{equation}
We have used eq.(2.10) together with
appropriate self-similarity conditions
corresponding to fractal dimensions $D_3$
and ${\tilde{D}}_3$ in configuration
and momentum space respectively. The coefficients $\alpha_{\mu \nu}$ in
eq.(3.6) are fixed by the transformation equations (3.5).
{}From eqs.(3.6) to (3.8) one easily obtains
stronger self-similarity constraints
as follows:
\begin{equation}
{\tilde{\rho}}_c(\lambda^2 y_i ', {\vec{q}}_{\perp i} \; ';
Y,{\vec{Q}}_{\perp})= \lambda^{({\tilde{D}}_3+D_3-4)(p-1)}
{\tilde{\rho}}_c(y_i ',{\vec{q}}_{\perp i} \; ';Y,{\vec{Q}}_{\perp})
\end{equation}
\begin{equation}
{\tilde{\rho}}_c(y_i ', \lambda^2 {\vec{q}}_{\perp i} \; ';
Y,{\vec{Q}}_{\perp})= \lambda^{({\tilde{D}}_3-D_3-2)(p-1)}
{\tilde{\rho}}_c(y_i ',{\vec{q}}_{\perp i} \; ';Y,{\vec{Q}}_{\perp})
\end{equation}
or in a unified form:
\begin{equation}
{\tilde{\rho}}_c(\lambda y_i ', \mu {\vec{q}}_{\perp i} \; ';
Y,{\vec{Q}}_{\perp})= \lambda^{{1 \over 2}({\tilde{D}}_3+D_3-4)(p-1)}
\mu^{{1 \over 2}({\tilde{D}}_3-D_3-2)(p-1)}
{\tilde{\rho}}_c(y_i ',{\vec{q}}_{\perp i} \; ';Y,{\vec{Q}}_{\perp})
\end{equation}
The geometrical interpretation of eq.(3.11) is that the fractal
${\tilde{F}}_3$ in $3D$ momentum space, generated by a fractal $F_3$
in configuration space, has necessarily the structure of a Cartesian
product of two fractals, ${\tilde{F}}_1$ and ${\tilde{F}}_2$, in rapidity
and transverse momentum space respectively, ${\tilde{F}}_3={\tilde{F}}_1
\times {\tilde{F}}_2$ \cite{falc:1}.
The fractal dimensions involved in this structure
are: ${\tilde{D}}_1={1 \over 2}(D_3+ {\tilde{D}}_3-2)$,
${\tilde{D}}_2={1 \over 2}({\tilde{D}}_3 -D_3 + 2)$ satisfying the
Cartesian product relation ${\tilde{D}}_1+{\tilde{D}}_2={\tilde{D}}_3$.
Similarly, in configuration space one obtains:
\begin{equation}
\rho_c(\lambda y_i ', \mu {\vec{r}}_{\perp i} \; ';
Y,{\vec{R}}_{\perp})= \lambda^{{1 \over 2}({\tilde{D}}_3+D_3-4)(p-1)}
\mu^{{1 \over 2}(D_3-{\tilde{D}}_3-2)(p-1)}
\rho_c(y_i ',{\vec{r}}_{\perp i} \; ';Y,{\vec{R}}_{\perp})
\end{equation}
showing that the fractal $F_3$ has also the structure of a Cartesian product,
$F_3=F_1 \times F_2$, of two fractals, one in rapidity $(F_1)$ and the other
in impact parameter space $(F_2)$. The corresponding fractal dimensions
in this case are: $D_1={1 \over 2}(D_3+ {\tilde{D}}_3-2)$,
$D_2={1 \over 2}(D_3-{\tilde{D}}_3 + 2)$ and $D_3=D_1+D_2$.
\par The realization of this geometry in terms of particle densities in
momentum space leads to the general form:
\begin{equation}
{\tilde{\rho}}_c(y_i ',{\vec{q}}_{\perp i} \; ';Y,{\vec{Q}}_{\perp})
=\rho_{c 1}(y_i ') {\tilde{\rho}}_{c 2}({\vec{q}}_{\perp i} \; ')
\Phi_{inv}^{(3)}(y ',{\vec{q}}_{\perp i} \; ';Y,{\vec{Q}}_{\perp})
\end{equation}
where $\rho_{c 1}$, ${\tilde{\rho}}_{c 2}$ obey self-similarity and
$\Phi^{(3)}_{inv}$ is invariant, under scale transformations~:
$y_i ' \rightarrow \lambda y_i '$, ${\vec{q}}_{\perp i} \rightarrow
\mu {\vec{q}}_{\perp i}$ as follows:
\begin{eqnarray}
\rho_{c 1}(\lambda y_i ') &=& \lambda^{(D_1-1)(p-1)}
\rho_{c 1}(y_i ') \nonumber \\
{\tilde{\rho}}_{c 2}(\mu {\vec{q}}_{\perp i} \; ') &=&
\mu^{({\tilde{D}}_2-2)(p-1)}
{\tilde{\rho}}_{c 2}({\vec{q}}_{\perp i} \; ') \nonumber \\
\Phi^{(3)}_{inv}(\lambda y_i ', \mu {\vec{q}}_{\perp i};
Y,{\vec{Q}}_{\perp})&=&
\Phi^{(3)}_{inv}( y_i ', {\vec{q}}_{\perp i};
Y,{\vec{Q}}_{\perp})
\end{eqnarray}
In summary, we have shown that the most general fractal solution of eq.(2.10)
in configuration and momentum space, associated to a second-order phase
transition during the development of the collision, has the structure of a
Cartesian product in both spaces. The geometrical characteristics of these
fractals are summarized as follows: (a) in configuration space,
$F_3=F_1 \times F_2$ with $D_3=D_1+D_2$ and (b) in momentum space,
${\tilde{F_3}}={\tilde{F}}_1 \times {\tilde{F}}_2$ with ${\tilde{F}}_1=
F_1$, ${\tilde{D}}_1=D_1$, ${\tilde{D}}_2=2-D_2$,
${\tilde{D}}_3=2+D_1-D_2$.
It is of interest to
note that the whole pattern of fractals in both spaces depends on two universal
indices $(D_1,D_2)$ which are constrained by the isothermal critical exponent
$\delta$, $D_1+D_2={3 \delta \over 1 + \delta}$.
\par In momentum space, the pattern of fractals has to be completed by
considering $1D$ projections of ${\tilde{F}}_2$ within its support space
(transverse momentum plane). Physically it is equivalent to search for $1D$
intermittency patterns in azimuthal angle $(\delta \phi)$ or transverse
momentum $(\delta q_{\perp})$ domains. According to deterministic fractal
geometry \cite{falc:1}
if ${\tilde{D}}_2 < 1$ the projection of ${\tilde{F}}_2$ onto
$1D$ subspaces of the transverse momentum plane is also a fractal
${\tilde{F}}_{\perp 1}$ with the same fractal dimension
${\tilde{D}}_{\perp 1}={\tilde{D}}_2$. Otherwise, if ${\tilde{D}}_2 > 1$, the
$1D$ projections of ${\tilde{F}}_2$ do not develop any fractal structure
(${\tilde{D}}_{\perp 1} = 1$). This property remains valid for random fractals
as well and can be verified for ${\tilde{F}}_2$ as follows: \\
Consider the projection of ${\tilde{F}}_2$ onto any $1D$ $u$-space by
integrating the densities ${\tilde{\rho}}_{c 2}({\vec{q}}_{\perp i} \; ')$
over the orthogonal $v$-space ($u$, $v$ are usual Cartesian coordinates in $2D$
transverse momentum). The projected densities are given by the integral:
\begin{equation}
{\tilde{\rho}}_{\perp c}^{(1)}(u_i)=
\int {\tilde{\rho}}_{c 2}(u_i,v_i) d v_1...d v_{p-1}
\end{equation}
which, after imposing the self-similarity constraint (3.14)
on ${\tilde{\rho}}_{c 2}$, gives:
\begin{equation}
{\tilde{\rho}}_{\perp c}^{(1)}(\lambda u_i)
=\lambda^{({\tilde{D}}_2-1)(p-1)}
{\tilde{\rho}}_{\perp c}^{(1)}(u_i)
\end{equation}
For ${\tilde{D}}_2 < 1$ eq.(3.16)
represents a self-similar structure in $1D$
$u$-space with fractal dimension
${\tilde{D}}_{\perp 1}={\tilde{D}}_2$ whereas
if ${\tilde{D}}_2 > 1$ the singularity for
$\lambda \rightarrow 0$ disappears
and the projected particle densities have a smooth non-fractal behaviour.
We therefore conclude that there are, in general, two possible classes of
fractal patterns in momentum space, depending on the actual value of the
fractal dimension ${\tilde{D}}_2$, which can be attributed to a critical
behaviour of the hadronic system at $T=T_c$. Although ${\tilde{D}}_2 > 1$
cannot be excluded on geometrical grounds alone, it can be easily seen
that it requires rather unrealistic values for the critical exponent $\delta$
and, therefore, can be safely ignored on physical grounds.
In fact, from the general relations  ${\tilde{D}}_2=2-D_3+D_1$ and
$D_3={3 \delta \over 1 + \delta}$ $(0<D_1<1)$, one obtains the bound
${\tilde{D}}_2 \leq {3 \over 1 + \delta}$, which in the sector
${\tilde{D}}_2 > 1$ implies $\delta < 2$. Since it is very unlikely for the
critical exponent $\delta$ to get such small values in any realistic $3D$
system undergoing a second-order phase transition \cite{stan:1} one may
consider
solution with ${\tilde{D}}_2 < 1$ as the only acceptable class of fractals in
momentum space associated to a second-order quark-hadron phase transition. The
geometrical characteristics (embedding dimensions, Cartesian products, fractal
dimensions $D_F$) of this solution are summarized as follows:
\begin{eqnarray}
3D &:& {\tilde{F}}_1(y) \times {\tilde{F}}_2(q_{\perp},\phi)
\; \; \; \; \; \; \; \; \; \; \; \; \; \; \; \; \; \;
\; \; \; \; \; \; \; \; \; \; \; \; \; \; \; \;
D_F=2+D_1-D_2 \nonumber \\
2D &:& {\tilde{F}}_2(q_{\perp},\phi)
\; \; \; \; \; \; \; \; \; \; \; \; \; \; \; \; \; \; \; \; \; \; \;
\; \; \; \; \; \; \; \; \; \; \; \; \; \; \; \; \; \; \; \; \; \; \;
\;
D_F=2-D_2 \nonumber \\
2D &:& {\tilde{F}}_{\perp 1}(q_{\perp}) \times {\tilde{F}}_1(y),
{\tilde{F}}_{\perp 1}(\phi) \times {\tilde{F}}_1(y) \; \; \; \; \;
\; \; \; \; \; \;  D_F=2+D_1-D_2 \nonumber \\
1D &:& {\tilde{F}}_{\perp 1}(q_{\perp}),
{\tilde{F}}_{\perp 1}(\phi)
\; \; \; \; \; \; \; \; \; \; \; \; \; \; \; \; \; \;
\; \; \; \; \; \; \; \; \; \; \; \; \; \; \; \; \; \;
D_F=2-D_2 \nonumber \\
1D &:& {\tilde{F}}_1(y)
\; \; \; \; \; \; \; \; \; \; \; \; \; \; \; \; \; \; \; \; \; \; \; \; \; \;
\; \; \; \; \; \; \; \; \; \; \; \; \; \; \; \; \; \; \; \; \; \; \; \;
\; \; D_F=D_1
\end{eqnarray}
The physical content of the geometrical pattern (3.17) may be revealed by
studying the behaviour of factorial moments in small domains of momentum
space, using appropriate self-similarity relations implied by the above
fractal structures. A set of power laws, dependent on two universal
parameters $(D_1,D_2)$, is expected to be generated, resulting in a
multidimensional intermittency pattern with a linear spectrum of indices.
The details of this treatment are given in section 4.
\begin{center}
\section{Multidimensional Intermittency}
\setcounter{equation}{0}
\end{center}
The scaled factorial moments, $F_p(\delta \Omega)$, in small domains of
momentum space \cite{bial:1}, are suitable for probing geometrical structures
of
form (3.17) and therefore connecting a second-order phase transition at the
level of quark-gluon dynamics with observable multiplicity fluctuations at
the level of the hadronic $S$-matrix. The basic ingredients in this approach
are the integrals:
\begin{equation}
F_p^{(3)} (\delta \Omega_3) \sim (\delta \Omega_3)^{-p} \int_{\delta \Omega_3}
{\tilde{\rho}}_c(y_i ', {\vec{q}}_{\perp i} \; '; Y, {\vec{Q}}_{\perp})
\{d y_i ' \} \{ d^2 {\vec{q}}_{\perp i} \; ' \} dY d^2 {\vec{Q}}_{\perp}
\end{equation}
$(p=2,3...)$ where integration is restricted in the vicinity of a fixed
configuration: $y_i '=0$, ${\vec{q}}_{\perp i} \; '=0$, $Y=Y_0 \; \;
{\vec{Q}}_{\perp}={\vec{Q}}_{\perp 0} \; \; \; \; (i=1,2...,p-1)$. In this
domain eq.(4.1) is simplified as follows:
\begin{equation}
F_p^{(3)} (\delta \Omega_3) \sim (\delta \Omega_3)^{1-p} \int_{\delta \Omega_3}
{\tilde{\rho}}_c(y_i ', {\vec{q}}_{\perp i} \; '; Y_0, {\vec{Q}}_{\perp 0})
\{d y_i ' \} \{ d^2 {\vec{q}}_{\perp i} \; ' \}
\end{equation}
The moments $F_p^{(3)} (\delta \Omega_3)$ refer to $3D$ intermittency but
similar quantities $F_p^{(1)}$ and $F_p^{(2)}$ in one or two dimensions are
easily obtained by projecting out onto appropriate momentum subspaces. Using
the
multiplicative form (3.13)  for the densities
${\tilde{\rho}}_c(y_i ', {\vec{q}}_{\perp i} \; '; Y_0, {\vec{Q}}_{\perp 0})$
we
may write a multidimensional pattern of moments as follows~:
\begin{equation}
F_p^{(1)} (\delta \Omega_1) \sim (\delta \Omega_1)^{1-p}  \int_{\delta
\Omega_1}
\rho_{c 1}(y_i ') \Phi_{inv}^{(1)}(y_i '; Y_0)
\{d y_i ' \}
\end{equation}
\begin{equation}
F_p^{(1)} (\delta \Omega_1 ') \sim
(\delta \Omega_1 ')^{1-p} \int_{\delta \Omega_1 '}
{\tilde{\rho}}_{\perp c}^{(1)}(u_i) \Phi_{inv}^{(1)}(u_i ; Q_u^{(0)})
\{d u_i  \}
\end{equation}
\begin{equation}
F_p^{(2)} (\delta \Omega_2 ') \sim
(\delta \Omega_2 ')^{1-p} \int_{\delta \Omega_2 '} \rho_{c 1}(y_i ')
{\tilde{\rho}}_{\perp c}^{(1)}(u_i)
\Phi_{inv}^{(2)}(y_i ',u_i ; Y_0,Q_u^{(0)})
\{d y_i ' \} \{d u_i  \}
\end{equation}
\begin{equation}
F_p^{(2)} (\delta \Omega_2 ) \sim
(\delta \Omega_2 )^{1-p} \int_{\delta \Omega_2 }
{\tilde{\rho}}_{c 2}({\vec{q}}_{\perp i} \; ')
\Phi_{inv}^{(2)}({\vec{q}}_{\perp i} \; ' ; {\vec{Q}}_{\perp 0})
\{d^2 {\vec{q}}_{\perp i} \; ' \}
\end{equation}
\begin{equation}
F_p^{(3)} (\delta \Omega_3 ) \sim
(\delta \Omega_3 )^{1-p} \int_{\delta \Omega_3 } \rho_{c 1}(y_i ')
{\tilde{\rho}}_{c 2}({\vec{q}}_{\perp i} \; ')
\Phi_{inv}^{(3)}(y_i ',{\vec{q}}_{\perp i} \; ' ; Y_0,{\vec{Q}}_{\perp 0})
\{d y_i ' \} \{d^2 {\vec{q}}_{\perp i} \; ' \}
\end{equation}
where $Q_u^{(0)}$ is the component of ${\vec{Q}}_{\perp 0}$ along the
$u$-direction in the transverse momentum plane. The integration volumes
$\delta \Omega_D$ are identified as follows:
$\delta \Omega_1 = \delta y \;,\; \delta \Omega_1 '= \delta q_{\perp}$
or $\delta \phi$, $\delta \Omega_2 ' =\delta q_{\perp} \delta y$ or
$\delta \phi \delta y$, $\delta \Omega_2 = \delta \phi \delta q_{\perp}$
and $\delta \Omega_3 = \delta y \delta \phi \delta q_{\perp}$. Their
size is taken
$\delta \Omega_D \sim M^{-D}$ with $M \gg 1$ and the dependence
of moments on $M$  reveals the intermittency effects in the process under
consideration.
\par
The integrands in eqs.(4.3) to (4.7)
 obey self-similarity corresponding to
codimensions $D-D_F=1-D_1,D_2-1,D_2-D_1,D_2, 1+D_2-D_1$. Therefore using
the transformations $y_i '=M^{-1} \theta_i, {\vec{q}}_{\perp i} \; '
=M^{-1} {\vec{s}}_i, u_i=M^{-1} \gamma_i$,
we obtain the power laws \cite{drem:1}:
\begin{eqnarray}
F_p^{(1)} (\delta \Omega_1) &\sim& M^{(1-D_1)(p-1)}	\int_{\delta \omega_1}
\rho_{c 1}(\theta_i) \Phi_{inv}^{(1)}(\theta_i; Y_0)
\{d \theta_i  \}
\nonumber \\
F_p^{(1)} (\delta \Omega_1 ') &\sim&
M^{(D_2-1)(p-1)} \int_{\delta \omega_1 '}
{\tilde{\rho}}_{\perp c}^{(1)}(\gamma_i) \Phi_{inv}^{(1)}(\gamma_i ; Q_u^{(0)})
\{d \gamma_i \}
\nonumber \\
F_p^{(2)} (\delta \Omega_2 ') &\sim&
M^{(D_2-D_1)(p-1)} \int_{\delta \omega_2 '} \rho_{c 1}(\theta_i)
{\tilde{\rho}}_{\perp c}^{(1)}(\gamma_i)
\Phi_{inv}^{(2)}(\theta_i ,\gamma_i ; Y_0,Q_u^{(0)})
\{d \theta_i \} \{d \gamma_i \}
\nonumber \\
F_p^{(2)} (\delta \Omega_2 ) &\sim&
M^{D_2(p-1)} \int_{\delta \omega_2 }
{\tilde{\rho}}_{c 2}({\vec{s}}_i)
\Phi_{inv}^{(2)}({\vec{s}}_i ; {\vec{Q}}_{\perp 0})
\{d^2 {\vec{s}}_i \}
\nonumber \\
F_p^{(3)} (\delta \Omega_3 ) &\sim&
M^{(1+D_2-D_1)(p-1)}  \nonumber\\
&\cdot& \int_{\delta \omega_3 }	\rho_{c 1}(\theta_i)
{\tilde{\rho}}_{c 2}({\vec{s}}_i)
\Phi_{inv}^{(3)}(\theta_i ,{\vec{s}}_i ; Y_0,{\vec{Q}}_{\perp 0})
\{d \theta_i  \}  \{d^2 {\vec{s}}_i \}
\end{eqnarray}
where the integration domains have finite, $M$-independent volumes
$\delta \omega_D \sim M^D \delta \Omega_D$. In summary, we may write the
complete multidimensional intermittency pattern corresponding to fractal
structures (3.17) as follows:
\begin{eqnarray}
\ln F_p^{(1)}(\delta y) &=& (1-D_1)(p-1) \ln M
\nonumber \\
\ln F_p^{(1)}(\delta \phi) &=& \ln F_p^{(1)}(\delta q_{\perp})=
(D_2-1)(p-1) \ln M
\nonumber \\
\ln F_p^{(2)}(\delta q_{\perp}, \delta y) &=&
\ln F_p^{(2)}(\delta \phi, \delta y)=
(D_2-D_1)(p-1) \ln M
\nonumber \\
\ln F_p^{(2)}(\delta \phi, \delta q_{\perp}) &=&
D_2(p-1) \ln M
\nonumber \\
\ln F_p^{(3)}(\delta y,\delta \phi, \delta q_{\perp}) &=&
(1+D_2-D_1)(p-1) \ln M
\end{eqnarray}
with the constraint $D_1+D_2={3 \delta \over 1 + \delta}$ $(\delta > 2)$.
A number of characteristic properties of the pattern (4.9) can be easily
observed, independently of the actual values of the universal indices
$D_1,D_2$: (a) There is a certain degree of degeneracy at the level of $1D$
and $2D$ moments, given by the equalities $F_p^{(1)}(\delta q_{\perp})=
F_p^{(1)}(\delta \phi)$ and $F_p^{(2)}(\delta q_{\perp}, \delta y)=
F_p^{(2)}(\delta \phi, \delta y)$ in the limit $M \gg 1$ ($\delta y \sim
M^{-1}, \delta q_{\perp} \sim M^{-1}, \delta \phi \sim M^{-1}$),
(b) introducing the index $\eta={D-D_F \over D}$ as a measure of intermittency
strength, eqs.(4.9) with $\delta > 2$ show that the weakest effect appears in
$1D$ intermittency in rapidity $(\eta=1 - D_1)$ and the strongest in $2D$
intermittency in transverse momentum plane $(\eta={D_2 \over 2})$,
and (c) the observed universality of $3D$ intermittency
index at present experim
\cite{fial:1} may be attributed to the universal character of $D_1,D_2$ and in
Fialkowski's notation, one may write $\phi_2={1 \over 3}(1+D_2-D_1)$.
\par In summary, we have found, under a certain number of simplified
assumptions, a complete multidimensional intermittency pattern dependent on two
universal indices $D_1,D_2$, which, we claim, is the most general pattern
associated to a second-order quark-hadron phase transition.
\begin{center}
\section{Phenomenology}
\setcounter{equation}{0}
\end{center}
At present experiments the energy density in hadron-hadron or nucleus-nucleus
collisions is not high enough to generate events of hadronized
quark-gluon plasma in abundance and therefore the universal pattern (4.9)
has to be modified by adding non-universal terms within the context of a
two-component model \cite{anto:1}. These terms represent conventional,
non-singu
correlations which contribute to the factorial moments as follows
\cite{anto:1}:
\begin{equation}
F_p^{(D)}(M)=\lambda_c c_p^{(D)} M^{(D-D_F)(p-1)}+G_p^{(D)}(M)
\end{equation}
where $G_p^{(D)}(M)$ are slowly varying functions of $M$, and
$\lambda_c \ll 1$ is the probability for producing quark-gluon plasma in the
actual experiment $(\lambda_c={\sigma_{q-g} \over \sigma_{in}})$. In a
simplified version of this model one may assume that the conventional
components $G_p^{(D)}(M)$ are constant ($M$-independent) terms $G_p^{(D)}$
and thus obtain a generalized Fialkowski's pattern as follows \cite{fial:1}:
\begin{equation}
\ln (F_p^{(D)} - G_p^{(D)})=\Bigl({D-D_F \over D}\Bigr)
(p-1) \ln(M^D) + \mbox{constant}
\end{equation}
valid for all moments $(p=2,3,...)$
and all dimensions $(D=1,2,3)$. Hence the
modified linear forms (5.2) are suitable for phenomenological comparison
with present experiments in order to reveal the universal indices
$\eta={D-D_F \over D}$ of the pattern (4.9). In the particular case of $3D$
intermittency, Fialkowski has verified the universality of the index $\eta$
by comparing the linear form (5.2) for $p=2,D=3$ with measurements
corresponding to a number of processes. In his notation ($\eta=\phi_2$ for
$p=2$ and $D=3$) the universal index
$\phi_2$ is fixed by the data in the range
$\phi_2=0.4-0.5$ \cite{fial:1}.
For phenomenological purposes we may now express the fractal dimensions
$(D_1,D_2)$ in terms of
the universal indices ($\phi_2,\delta$), using the relations:
$D_1+D_2={3 \delta \over 1 + \delta}, \; \; \;
\phi_2={1 \over 3}(D_2-D_1+1)$.
We find:
\begin{equation}
D_1=-{3 \phi_2 \over 2}+{1 \over 2}+{3 \delta \over 2(1 + \delta)}
\; \; \; \; \; \; \; \; D_2={3 \phi_2 \over 2}-{1 \over 2}+
{3 \delta \over 2(1 + \delta)}
\end{equation}
Equations (5.3) together with the constraints $0<D_1<1$, $1<D_2<2$,
$\delta>2$ put a restriction on the index $\phi_2$:
\begin{equation}
-{1 \over 3}+{ \delta \over 1 + \delta} \; < \; \phi_2 \; < \;
{5 \over 3}-{ \delta \over 1 + \delta}
\end{equation}
leading to a maximal range of values allowed for $\phi_2$, ${1 \over 3}
< \phi_2 < 1$. The lower bound gives a measure of $3D$ intermittency
strength and it is remarkable that the phenomenological values of $\phi_2$
lie in this domain, close to the lower bound \cite{fial:1}.
The bounds (5.4) on the other hand show a qualitative interrelation of the
measurable index $\phi_2$ with the critical exponent $\delta$ and may lead
to a phenomenological estimate of the allowed values for this fundamental
parameter. In fact,
 for two distinct cases (a) $\delta=3$ corresponding to
the mean-field approximation and
(b) $\delta \approx 5$ corresponding to the
universality class of the $n=4$ Heisenberg magnet
($M \sim H^{1 \over \delta}$) suggested by Wilczek \cite{wilz:1},
we have (a)
${5 \over 12} < \phi_2 < {11 \over 12}$ and (b) ${1 \over 2} < \phi_2 <
{5 \over 6}$ respectively. We observe that Fialkowski's phenomenological
estimate of $\phi_2$ is consistent with the
mean-field approximation value
$\delta=3$ but marginally consistent with the $n=4$ magnet value $\delta
\approx 5$. It seems that present data favours rather small
values of the critical exponent $\delta \leq 4$, close to the mean-field
approximation value $\delta=3$ [Fig.1].
\begin{figure}[ht]
\begin{center}
\mbox{}
%\mbox{\psfig{file=anf1.eps.b,width=12cm,height=12cm}}
\caption[.]{The bounds eq.(5.4) in a $\delta$-$\phi_2$
plot together with
Fialkowski's values for $\phi_2$ (shaded region).}
\end{center}
\end{figure}
For completeness we may write similar bounds for $1D$ and $2D$ intermittency
indices, $\eta={D-D_F \over D}$. For this purpose and according to the
pattern (4.9) we put $\eta_1=1-D_1$ ($\delta y$), $\eta_{1\perp}=D_2-1$
($\delta \phi$ or $\delta q_{\perp}$) for $1D$ intermittency and $\eta_2
={D_2-D_1 \over 2}$ ($\delta y \delta q_{\perp}$ or $\delta y \delta \phi$)
$\eta_{2 \perp}={D_2 \over 2}$  ($\delta q_{\perp} \delta \phi$) for $2D$
intermittency. Following the same procedure as in the case of $3D$
intermittency we find, for a given value of the critical exponent $\delta$,
the bounds
\begin{eqnarray}
1D &:& 0 \; < \; \eta_1 \; < \; 3 - {3 \delta \over 1 + \delta}
\; \; \; ; \; \; \; \; {3 \delta \over 1 + \delta} -2 \; < \;
\eta_{1 \perp} \;<\; 1 \nonumber \\
2D &:& {3 \delta \over 2(1 + \delta)}-1 \; < \; \eta_2 \; < \; 2 -
{3 \delta \over 2(1 + \delta)} \;  ; \; \;
{3 \delta \over 2(1 + \delta)} - {1 \over 2} \; < \; \eta_{2 \perp} \;
< \; 1
\end{eqnarray}
showing a strong $2D$ intermittency effect in the
transverse momentum plane
$({1 \over 2}  <  \eta_{2 \perp}  <	1)$.
Finally, the phenomenological capacity of our approach is illustrated in
Fig.2 considering as a basic ingredient Fialkowski's universality for $3D$
intermittency at the level of the
second factorial moment $F_2^{(3)}(M)$. In
Fig. 2a the
consistency of the bounds (5.4) with the trend of the data is shown
and the accumulation tendency towards the lower bound $\phi_2={1 \over 3}$
is illustrated. We have used the outcome of a
recent analysis of Fialkowski,
based on NA35 measurements \cite{bach:1} related to a number of processes
$(OAu, SS, SAu)$. In Fig. 2b a particular multidimensional
intermittency pattern is singled-out corresponding to the choice $\delta=3$
(mean-field approximation) and $\phi_2=0.44$ (average value from
Fialkowski's analysis) in the general solution (4.9). This illustration
suggests that a phenomenological exploitation of our model would require
an extension of Fialkowski's analysis taking into consideration measurements
of $1D$ and $2D$ patterns as well. In such a treatment one could be able
not only to reveal a universal multidimensional intermittency effect
associated to a second-order quark-hadron phase transition, but also to
measure the critical exponent $\delta$ since, for a given value of $\phi_2$,
the slopes of $1D$ and $2D$ moments in Fig.2b are sensitive to the actual
value of $\delta$.
\begin{figure}[ht]
\begin{center}
\mbox{}
%\mbox{\psfig{file=anf2.eps.b,width=12cm,height=12cm}}
\caption[.]{
(a) The moments (5.2) for $p=2$,
$D=3$ corresponding to the limiting
slopes $\phi_2=1/3, 1$ and (b) a particular multidimensional
intermittency
pattern eq.(4.9) with $\delta=3$, $\phi_2=0.44$, are compared with
Fialkowski's data analysis.
The power laws are arbitrarily normalized at the
same point.}
\end{center}
\end{figure}
\begin{center}
\section{Conclusions}
\setcounter{equation}{0}
\end{center}
In this work we have studied the general structure of multidimensional
intermittency patterns in hadronic collisions generated by a second-order
quark-hadron phase transition. For this purpose we have searched for the
most general class of fractals in momentum space associated to
self-similar structures in configuration space normally developed at
$T=T_c$ \cite{stin:1}.
The significance of self-similarity in configuration s   pace
(space-time) for intermittency phenomena has been recently emphasized by
Bialas in connection with the Bose-Einstein
interference effect \cite{bial:2}.A basic
ingredient in our approach is the standard inside-outside cascade picture
for the development of the collision, in the hadronization sector,
allowing the
use of rapidity as a longitudinal coordinate both in configuration and
momentum space $(\xi \approx y)$. In this framework, employing the
reduction formalism for the inclusive $S$-matrix, an integral
representation connecting hadronic densities in momentum and configuration
space was established. We have found that the general fractal solution both
in momentum and configuration space has the geometrical structure of a
Cartesian product of two fractals, one in rapidity space and the other in
the
transverse-momentum or impact-parameter space respectively. The resulting
multidimensional intermittency pattern has a linear spectrum of indices,
as expected, dependent on two universal parameters $(D_1,D_2)$,
geometrically identified with the fractal dimensions in rapidity $(D_1)$
and impact parameter space $(D_2)$. Physically these universal indices are
constrained by the isothermal critical exponent $\delta$ of the strongly
interacting system at $T=T_c$, $D_1+D_2={3 \delta \over 1 + \delta}$. For
realistic values $\delta > 2$ we have seen that $1D$ projections of the
fractal system in the
transverse-momentum plane are also fractals with fractal
dimension ${\tilde{D}}_{\perp 1}=2-D_2$. Hence $1D$ intermittency in
azimuthal angle $(\delta \phi)$ or transverse momentum $(\delta q_{\perp})$
reflects the development of a $2D$ fractal structure in transverse momentum
plane with fractal dimension ${\tilde{D}}_2={\tilde{D}}_{\perp 1} \;
< \; 1$. In this case $1D$ intermittency (in $\delta \phi$ or
$\delta q_{\perp}$) is weaker than the corresponding effect in $2D$
($\delta \phi \delta q_{\perp}$) in accordance with previous proposals
about the dependence of intermittency strength on dimensionality in
momentum space \cite{ochs:1}. On the contrary, $1D$ intermittency in rapidity
is
genuine effect since the fractal structure ${\tilde{F}}_1$ in the Cartesian
product ${\tilde{F}}={\tilde{F}}_1 \times {\tilde{F}}_2$ eq.(3.17) is an
intrinsic property of the critical system and not the artefact of a
structure in higher dimensions. Its strength, in particular, depends on
the basic index $D_1$ and is not generated by projecting out fractals
embedded in higher dimensions.
\par Our approach to intermittency, based on the critical properties of the
strongly interacting system, is suitable for hadronic collisions $(hh)$
and especially for relativistic heavy-ion processes $(AA)$ with high
energy-density.
It is complementary to perturbative QCD-inspired treatments,
mainly applicable to $e^+ e^-$ intermittency where a branching mechanism at
the parton level is expected to dominate the density fluctuations of the
produced hadrons \cite{ochs:2}.
\par The phenomenological implications of our solution eq.(4.9) can only
be revealed after subtracting the conventional contributions to factorial
moments in the framework of a two-component model \cite{anto:1}. A simplified
version of this treatment leads naturally to Fialkowski's universality in
$3D$; in order to derive a direct connection of measurable
intermittency indices with the critical exponent $\delta$ at $T=T_c$ we
have argued that an extension of Fialkowski's analysis in one and two
dimensions is needed.
\par Finally, it is of interest to note that a natural extension of our
approach from static to dynamic scaling hypothesis could easily
accommodate non-trivial modifications of the intermittency pattern
eq.(4.9),
at least at the level of second moments $F_2^{(D)}$, due to the presence
of the dynamic critical exponent \cite{anto:2}. It is tempting to associate
such modifications with a more realistic picture of inside-outside
hadronization process as a critical phenomenon out of equilibrium
\cite{anto:2,curr:2}
\par
In conclusion, we have shown that multi\-dimensional intermittency
phenomena in hadronic collisions may originate from a simple pattern of
fractal structures
(mo\-no\-frac\-tals) in configuration space (space-time)
characteristic of a second-order phase transition. A careful study of these
phenomena is therefore necessary, both theoretically and experimentally, in
an attempt to establish a link between criticality of the underlying
theory (QCD at $T=T_c$) and complexity of the hadronic system at the
transition temperature ($S$-matrix at $T=T_c$).

\end{document}